# A µSR study of the magnetoresistive ruthenocuprates $RuSr_2Nd_{1.8-x}Y_{0.2}Ce_xCu_2O_{10-\delta}$ (x = 0.95 and 0.80)

A. C. Mclaughlin [1]*, J. P. Attfield [2], J. Van Duijn [3] and A. D. Hillier [4]

[1] Department of Chemistry, University of Aberdeen, Meston Walk, Aberdeen AB24 3UE, UK.

[2] Centre for Science at Extreme Conditions and School of Chemistry, University of Edinburgh, King's Buildings, Mayfield Road, Edinburgh EH9 3JZ.

[3] Instituto de Energías Renovables, Universidad de Castilla la Mancha, Albacete, E02006, Spain.

[4] ISIS facility, Rutherford Appleton Laboratory, Chilton, Didcot, Oxfordshire, OX11 0QX, UK.

Corresponding author: A. C. Mclaughlin (a.c.mclaughlin@abdn.ac.uk)

**Abstract**

Zero field muon spin relaxation (ZF-µSR) has been used to study the magnetic properties of the underdoped giant magnetoresistive ruthenocuprates (GMR) $RuSr_2Nd_{1.8-x}Y_{0.2}Ce_xCu_2O_{10-\delta}$ (x = 0.95, 0.80). The results show a gradual loss of initial asymmetry $A_0$ at the ruthenium spin transition temperature, $T_{Ru}$. At the same time the electronic relaxation rate, λ shows a gradual increase with decreasing temperature below $T_{Ru}$. These results have been interpreted as evidence for Cu spin cluster formation below $T_{Ru}$. These magnetically ordered clusters grow as the temperature is decreased thus causing the initial asymmetry to decrease slowly. GMR is observed over a wide temperature range in the materials studied and the magnitude increases as the temperature is reduced from $T_{Ru}$ to 4 K which suggests a relation between Cu spin cluster size and |-MR|.

## Introduction

The $RuSr_2RCu_2O_8$ (R=rare earth, Ru-1212) [1, 2, 3, 4] and $RuSr_2R_{2-x}Ce_xCu_2O_{10-\delta}$ (Ru-1222) [1, 2, 5] type ruthenocuprates have been well studied in recent years due to the observation of coexisting weak ferromagnetism (W-FM) and superconductivity. W-FM is observed in the ruthenate layer below 100 – 150 K and superconductivity in copper oxide planes below $T_c \sim 50$ K. Neutron scattering experiments on $RuSr_2GdCu_2O_8$ [6] and $Pb_2RuSr_2Cu_2O_8Cl$ [7] have recently shown that the Ru spins order in a G-type antiferromagnetic (AFM) arrangement in which the spins are aligned along *c* and are coupled antiferromagnetically in all three crystallographic directions. An upper limit of 0.1 $\mu_B$ was obtained for the ferromagnetic component. Upon application of a magnetic field the Ru spins cant further away from the G-type magnetic structure and at 7 T the order of the Ru spins is almost fully ferromagnetic. The magnetic structure of superconducting $RuSr_2Y_{1.5}Ce_{0.5}Cu_2O_{10-\delta}$ ($T_c = 35$ K) has recently been reported [8]. Neutron diffraction studies have shown that below the Ru spin ordering temperature-$T_{Ru}$, neighbouring spins align in an antiparallel arrangement in the *ab* plane while both antiferromagnetic and ferromagnetic alignment along *c* is observed [8].

The under-doped $RuSr_2Nd_{1.8-x}Y_{0.2}Ce_xCu_2O_{10-\delta}$ ($0.7 \leq x \leq 0.95$) ruthenocuprates are not superconducting but exhibit large negative magnetoresistances ((MR = (($\rho_H$-$\rho_0$)/$\rho_0$))) at low temperature up to -49% at 4 K in a 9 T field [9, 10, 11 12] demonstrating strong spin-charge coupling within the $CuO_2$ planes. -MR initially rises to ~2% at $T_{Ru}$, as observed in other superconducting ruthenocuprates but increases dramatically on further cooling. Neutron diffraction studies have shown that the Ru spins order in the same antiferromagnetic arrangement as for the superconducting analogue [8, 9]. In the underdoped material the Cu spins appear to order antiferromagnetically in the *ab* plane with a (½ ½ 0) superstructure below a second transition at $T_{Cu}$. Upon application of a magnetic field the Ru and Cu moments cant into a ferromagnetic alignment in the *ab* plane. The variation of –MR with temperature and field are characteristic of charge transport by magnetopolarons – small ferromagnetic regions surrounding each Cu-hole within a matrix of

3antiferromagnetically ordered $Cu^{2+}$ spins [13]. An applied magnetic field cants the Ru spins into a ferromagnetic arrangement, which induces partial ferromagnetism in the $CuO_2$ planes thereby increasing the mobility of the magnetopolarons, giving the observed, negative MR's [9, 10]. The magnetotransport in the Ru-1222 ruthenocuprates is also very sensitive to lattice effects. In a series of $RuSr_2R_{1.1}Ce_{0.9}Cu_2O_{10-\delta}$ (R = Nd, Sm, Eu, and Gd with Y) samples where the hole doping level is constant, the high field MR does not correlate with the paramagnetic moment of the $R$ cations, but shows an unprecedented crossover from negative to positive MR values as $<r_A>$, the mean A site ($R_{1.1}Ce_{0.9}$) cation radius decreases [10].

XANES studies have shown that Ru remains in the formal +5 state in the 1222 ruthenocuprates (although this is not true of 1212 types), e.g. the measured Ru valence remains at 4.95(5) as $x$ increases from 0.5 to 1.0 in $RuSr_2Gd_{2-x}Ce_xCu_2O_{10-\delta}$ [14], and so reliable Cu hole doping concentrations $p$ can be calculated from the cation and oxygen contents as $p = (1 - x - 2\delta)/2$. A recent neutron diffraction study of $RuSr_2Nd_{2-x-y}Y_yCe_xCu_2O_{10-\delta}$ has shown that separate Ru and Cu spin ordering transitions are observed, with spontaneous Cu antiferromagnetic order for low hole doping levels $p < 0.02$ [15]. However it was shown that the Cu spin order in the $0.02 < p < 0.06$ pseudogap region is induced by the Ru spin order. The thermal evolution of the (½ ½ 2) magnetic intensity (associated with the ordering of the Cu spins) obtained from variable temperature neutron diffraction data for $RuSr_2Nd_{1.0}Y_{0.1}Ce_{0.9}Cu_2O_{10-\delta}$ (p = 0.017) can be described by the critical expression $I/I_{4K}= (1-T/T_c)^{2\beta}$; and a fit in the $T_c/2 < T < T_c$ regions give $\beta = 0.38$ [15] which is typical of a three dimensionally ordered antiferromagnet. In contrast the thermal evolution of the (½ ½ 2) intensity for both the p = 0.033 and p = 0.055 samples is different to that for p = 0.017, as the intensity rises gradually below an ill-defined transition, and is not saturated down to 4 K. This is characteristic of an induced (non-spontaneous) magnetic order.

In order to further investigate this induced Cu magnetic order, variable temperature muon spin relaxation experiments on two Ru-1222 samples namely $RuSr_2Nd_{1.8-x}Y_{0.2}Ce_xCu_2O_{10-\delta}$ with x = 0.95 and 0.80 and corresponding $p$= 0.021 and 0.046 have been performed. Both samples are in the induced antiferromagnetic region of the 1222 ruthenocuprate electronic phase diagram [15]. The results show evidence of Cu spin cluster formation below $T_{Ru}$.

## Experimental

Samples of $RuSr_2Nd_{1.8-x}Y_{0.2}Ce_xCu_2O_{10-\delta}$ (x = 0.95 ($p$ = 0.021) and x = 0.8 ($p$ = 0.046)) were prepared by the solid-state reaction of stoichiometric powders of $Nd_2O_3$, $Y_2O_3$, $RuO_2$, CuO, $CeO_2$ and $SrCO_3$, as described elsewhere [10]. The μSR study was carried out using the MuSR spectrometer, in longitudinal geometry, at the ISIS pulsed muon and neutron facility. The powdered sample was mounted onto a silver plate in a CCR cryostat and cooled down to a base temperature of 13 K. To measure the time evolution of the muon spin polarisation, emitted decay positrons were collected in the forward (F) and backward (B) detector arrays relative to the initial muon spin direction. The muon asymmetry data $G_z(t)$ is determined from the F and B positron counts $N_{F/B}(t)$ using the equation

$$G_z(t) = \frac{N_F(t) - \alpha N_B(t)}{N_F(t) + \alpha N_B(t)}. \qquad (1)$$

The μSR spectra were collected at various temperatures upon warming to 300 K in zero field. Magnetic susceptibility data were collected between 5 K and 300 K on a Quantum Design SQUID magnetometer in an applied field of 100 Oe after zero-field (ZFC) and field cooling (FC).

## Results and Discussion

The temperature dependence of the magnetic susceptibility of $RuSr_2Nd_{1.8-x}Y_{0.2}Ce_xCu_2O_{10-\delta}$ (x = 0.95, 0.8) are displayed in Fig. 1. It has previously been shown that $T_{Ru}$ can be determined by extrapolating the maximum (-dM/dT) slope to zero magnetization, while $T_{Cu}$ is estimated from the temperature of the maximum zero field cooled magnetization ($M_{max}$) [15]. The magnetic transitions determined in this way correlate well with transitions established from neutron diffraction. $T_{Ru}$ and $T_{Cu}$ are estimated at 150 K and 73 K and 95 K and 45 K for x = 0.95 and x = 0.8 respectively. Previous results have shown that large negative magnetoresistances are observed below $T_{Ru}$ for both samples ($MR_{7T}$ (5 K) = -20% and -22% for x = 0.95 and 0.80 respectively) [10, 15].

Figure 2 shows representative zero field (ZF) μSR asymmetry, $G_z(t)$, for $RuSr_2Nd_{1.0}Y_{02}Ce_{0.8}Cu_2O_{10-\delta}$ at temperatures between 25 K, 50 K and 300 K. The μSR spectra, for all temperatures and each composition, are best described using the function

$$G_z(t) = A_0\left(\tfrac{1}{3} + \tfrac{2}{3}\left(1 - \sigma^2 t^2\right)\exp\left(-\tfrac{\sigma^2 t^2}{2}\right)\right)\exp(-\lambda t) + A_{bck} \quad (2)$$

where $A_0$ is the initial asymmetry, $\sigma$ is the nuclear moment contribution, $\lambda$ is the electronic relaxation rate and $A_{bck}$ is a time independent background coming from those muons which are implanted into the exposed Ag around the sample. Equation 2 can be broken down into two components. Firstly, a contribution associated with nuclear moments, known as a Gaussian Kubo Toyabe function [16, 17].

$$\left(\frac{1}{3} + \frac{2}{3}\left(1 - \sigma^2 t^2\right)\exp\left(-\frac{\sigma^2 t^2}{2}\right)\right), \quad (3)$$

and secondly a contribution associated with electronic moments,

$$\exp(-\lambda t). \quad (4)$$

These processes are independent and therefore can be added together multiplicatively. Each of these contributions have been considered in turn. The nuclear contribution in Eq. 2 is derived by assuming that the muons are stationary (i.e. not hopping within the lattice) and that the three orthogonal components of the magnetic field have a Gaussian distribution with a zero mean value and a root mean squared width, $\Delta/\gamma_\mu$ [16, 17]. The nuclear depolarisation rate, $\sigma$, is related to $\Delta$ by the relation $\sigma^2 = \gamma_\mu^2 \Delta^2$, where $\gamma_\mu$ is the gyromagnetic ratio of the muon ($\gamma_\mu$ = 13.55kHz/G). Above the Ru spin ordering temperature the nuclear depolarisation rate, $\sigma$, for $RuSr_2Nd_{1.8-x}Y_{0.2}Ce_xCu_2O_{10-\delta}$ is equal to 0.060(1) μs$^{-1}$ and 0.058(1) μs$^{-1}$ for x = 0.8 and x = 0.95 respectively. If long range magnetic order is present then the muon relaxation data can generally be fit to as oscillating function with frequency proportional to the internal field of the sample. However, if the frequency of precession is outside the time window accessible by the MuSR spectrometer, which is approximately 10 MHz, then the resultant muon beam will be depolarised which causes a loss of the initial asymmetry to 1/3 of the high temperature value. For a system of non interacting electronic spins the muon spin relaxation function generally takes the form

$$G_z(t) = \exp(-\lambda t) \qquad (5)$$

which is characterised by a unique spin relaxation rate. The exponential decay of the atomic muon spin relaxation corresponds to a simple exponential form of the time dependent autocorrelation function at the muon site. $\lambda$ is the electronic contribution to the relaxation rate and is related to the second moment of the atomic magnetic field distribution, $<\Delta B>^2$, and the correlation time, $\tau_c$ by the relation $\lambda = \gamma_\mu^2 \langle 2\Delta B^2 \rangle \tau_c$. The temperature dependence of the magnetic depolarisation rate will therefore vary as the internal field and the correlation time of the spin fluctuations change. Although, one might expect that for $RuSr_2Nd_{1.8-x}Y_{0.2}Ce_xCu_2O_{10-\delta}$ with its three magnetic ions that we would observe independent relaxation rates it can be seen from Fig. 2 that the muon depolarisation spectra are well described by a single exponential decay.

The µSR spectra from the $RuSr_2Nd_{1.8-x}Y_{0.2}Ce_xCu_2O_{10-\delta}$ samples show no evidence of oscillations, but do display a loss of initial asymmetry, which suggests that the frequency of precession is outside the time window accessible by the MuSR spectrometer; the electronic contribution can be well described by Equation 2. Figure 2 shows fits to the zero field data for $RuSr_2Nd_{1.0}Ce_{0.8}Y_{0.2}Cu_2O_{10-\delta}$ for selected temperatures. An excellent fit to the data is obtained at all temperatures for both samples, demonstrating that $\sigma$ is temperature independent and confirming that the muon is static within the lattice. If this were not the case, $\sigma$ would decrease as the temperature increases as a result of motional narrowing. $A_{bck}$ is also temperature independent for both samples at all temperatures. For both samples a gradual loss of asymmetry is observed below $T_{Ru}$, with a steeper loss of asymmetry at lower temperature which coincidences with the order temperature of the Cu spins( ~73K for x = 0.95 and ~ 45 K for x = 0.80) (Figure 3). The temperature dependence of $\lambda$ is shown in Figure 4 and evidences a gradual increase in $\lambda$ with decreasing temperature below $T_{Ru}$. Upon further cooling, $\lambda$ continues to increase which is a consequence of the decrease in the magnetic fluctuations of the $Nd^{3+}$ spins. In interpreting the results we will consider both the temperature dependence of the initial asymmetry and lambda together.

If a bulk magnetic transition, with an internal field greater than the frequency response of the MuSR spectrometer, is observed we would expect a drop in $A_0$ to 1/3 of the high temperature value, along with a simultaneous peak in $\lambda$ (as a result of a slowing down of the Ru moments at $T_{Ru}$). However this is not observed experimentally for $RuSr_2Nd_{1.8-x}Y_{0.2}Ce_xCu_2O_{10-\delta}$ (x = 0.95 and 0.80) where an unusual gradual increase in $\lambda$ (and decrease in $A_0$) is observed below $T_{Ru}$. This implies that there are regions within the sample in which there is a high internal field, due to static magnetic order and regions that are still paramagnetic. This is difficult to reconcile with the magnetization and neutron diffraction data, where clear magnetic transitions are evidenced for both Ru and Cu spins. The absence of a conventional transition at $T_{Ru}$ contrasts with results from previous variable temperature μSR experiments on superconducting 1212 and 1222 ruthenocuprates [3, 18] and suggests that the muon could be at a site of high symmetry in $RuSr_2Nd_{1.8-x}Y_{0.2}Ce_xCu_2O_{10-\delta}$. In order to determine if this is possible a finite element analysis has been performed, over a 4x4x4 unit cell (in order to reduce end effects) to determine σ using the relation

$$\sigma^2 = \frac{1}{6}I(I+1)\frac{8}{3}\left(\frac{\mu_0 \hbar}{4\pi}\gamma_\mu \gamma_n\right)^2 \left(1 + \frac{3}{8}\frac{I+\frac{1}{2}}{I(I+1)}\right)\sum_{i=1}^{N}\frac{1}{r_i^6} \qquad (6)$$

where I is the spin and $\gamma_n$ is the gyromagnetic ratio of the nucleus (for more details see reference 19). Results show that there are four planes in which the muon could locate, which correspond to the oxygen layers as shown in Figure 5. If we relate the Ru magnetic structure obtained from neutron diffraction [9, 15] and the site of the muon determined from finite element analysis, it appears highly likely that the muon is at a site of high magnetic symmetry; similar sites are found in other oxides [20]. Hence the field at the muon site as the Ru spins order should be zero and therefore there should be no change in the asymmetry, $A_0$, below $T_{Ru}$. This suggests that the change in $A_0$ below $T_{Ru}$ therefore is a result of the field from the Cu spins. Previous neutron diffraction measurements have shown that antiferromagnetic order of the Cu spins is induced by the W-FM of the Ru spins in the $RuSr_2Nd_{1.8-x}Y_{0.2}Ce_xCu_2O_{10-\delta}$ ruthenocuprates [9, 15]. We suggest that ordering of the Ru spins actually results in Cu cluster formation below $T_{Ru}$ i.e. regions in which the sample has regions of ordered Cu spins and regions where the Cu spins are non magnetic. Indeed, the gradual reduction in the initial asymmetry would support this theory (Fig. 3). As part of the sample orders the internal field causes a loss of asymmetry. These magnetically ordered clusters grow as the temperature is decreased thus causing the initial asymmetry to decrease slowly. Finally bulk antiferromagnetic order of the Cu spins will be observed at low temperature as observed by neutron diffraction. The temperature dependence of $\lambda$ can also be explained by cluster formation. As the temperature decreases towards $T_{Ru}$, $\lambda$ increases which shows that there is a decrease in the spin fluctuation rate as expected. However the critical divergence in $\lambda$ which is normally observed at a magnetic phase transition, and has been previously been observed at $T_{Ru}$ in superconducting ruthenocuprates [3, 18], is not evidenced and instead $\lambda$ increases gradually with decreasing temperature. Hence a gradual reduction in the magnetic fluctuation rate is observed as the temperature is reduced and the magnetically ordered Cu clusters grow. At low temperature the relaxation is dominated by the fluctuation of the $Nd^{3+}$ moments. Magnetic scattering from the Cu spin order will only be observed from neutron diffraction when the clusters reach a critical size which explains the distinct Cu magnetic ordering temperatures previously reported for $RuSr_2Nd_{1.8-x}Y_{0.2}Ce_xCu_2O_{10-\delta}$ [9, 15]. In corroboration similar cluster growth has been evidenced from µSR experiments on CMR materials $Sr_2RMn_2O_7$ [21] and is also well documented in cuprates [22].

Previous work has shown that there is no evidence of a magnetostriction in the Ru-O bond lengths or angles but surprisingly an anomaly is apparent when the contributions of the interplanar Cu-Cu distance $d_I$ (Fig. 5) and the thickness of the ruthenocuprate slab $d_{RC}$ to the c-axis length are considered [15]. A clear change in slope of $d_I$ at $T_{Ru}$ has been observed for several samples. By contrast, no anomalies are observed in the thermal variation of $d_{RC.}$ The origin of this effect was not previously understood, but these results suggest that this magnetostriction arises as a result of the cluster formation of Cu spins below $T_{Ru}$. Intriguingly a similar anomaly has previously been observed at the Ru spin ordering temperature from neutron diffraction measurements on superconducting RuSr$_2$GdCu$_2$O$_8$ [23]. This would suggest that Cu spin cluster formation arises below T$_{Ru}$ in both superconducting and non-superconducting ruthenocuprates. If this is the case it would be vital to determine how the Cu spin clusters transform as superconductivity is approached. Hence further investigation of the Cu spin dynamics in superconducting ruthenocuprates below T$_{Ru}$ are warranted.

In conclusion we have recorded variable temperature zero field μSR data for two magnetoresistive underdoped 1222 ruthenocuprates RuSr$_2$Nd$_{1.8-x}$Y$_{0.2}$Ce$_x$Cu$_2$O$_{10-\delta}$ (x = 0.95 and 0.80). An excellent fit of the asymmetry data to Equation (2) is obtained at all temperatures. Results demonstrate an unusual variation of both initial asymmetry, A$_0$ and relaxation rate λ with temperature which reveals that order of the Ru spins induces Cu spin clusters below T$_{Ru}$ which grow in size as the temperature is reduced. The magnitude of the –MR in the RuSr$_2$Nd$_{1.8-x}$Y$_{0.2}$Ce$_x$Cu$_2$O$_{10-\delta}$ underdoped ruthenocuprates is observed to increase below T$_{Ru}$ reaching maximum values at the lowest temperatures [9, 15] suggesting a relation between Cu spin cluster size and |-MR|.

**Acknowledgements**

ACM thanks the Leverhulme Trust and EPSRC for financial support and STFC for access to the MuSR spectrometer at ISIS.

41

Figure Captions

FIG. 1  Variable temperature magnetization data (zero-field and field cooled) for $RuSr_2Nd_{1.8-x}Ce_xY_{0.2}Cu_2O_{10-\delta}$ (x = 0.95, 0.80) solid solutions recorded in H = 100 Oe.

FIG. 2  (Color online) Typical μSR spectra for $RuSr_2Nd_{1.0}Ce_{0.8}Y_{0.2}Cu_2O_{10-\delta}$ in zero field at selected temperatures between 20 and 300 K. The solid line shows the fit of the data to equation 1.

FIG. 3  (Color online) Temperature dependence of the initial asymmetry $A_0$ for $RuSr_2Nd_{0.85}Ce_{0.95}Y_{0.2}Cu_2O_{10}$ and $RuSr_2Nd_{1.0}Ce_{0.8}Y_{0.2}Cu_2O_{10}$ evidencing a gradual reduction in asymmetry at $T_{Ru}$.

FIG. 4  (Color online) Variation of the electronic relaxation rate, λ with temperature for $RuSr_2Nd_{0.85}Ce_{0.95}Y_{0.2}Cu_2O_{10}$ and $RuSr_2Nd_{1.0}Ce_{0.8}Y_{0.2}Cu_2O_{10}$.

FIG. 5  Crystal structure of $RuSr_2Nd_{1.8-x}Ce_xY_{0.2}Cu_2O_{10-\delta}$. The oxygen plane in which the muons are located is arrowed and the interplanar separation of $CuO_2$ planes, $d_I$ is also labelled.

**Fig 1**

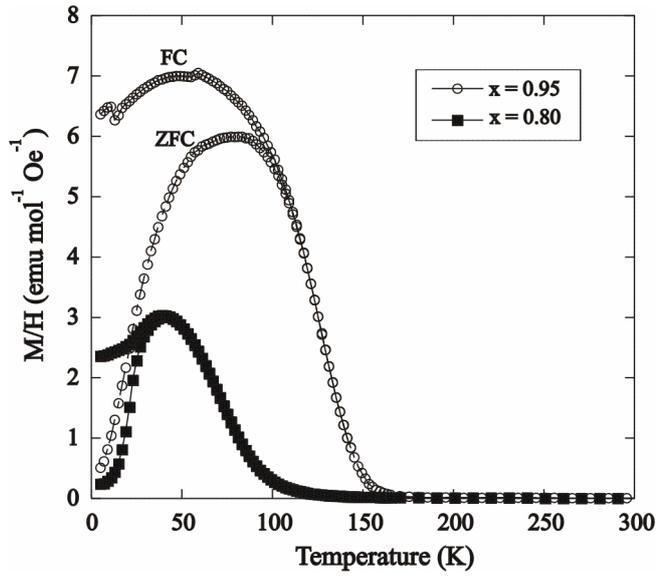

**Fig 2**

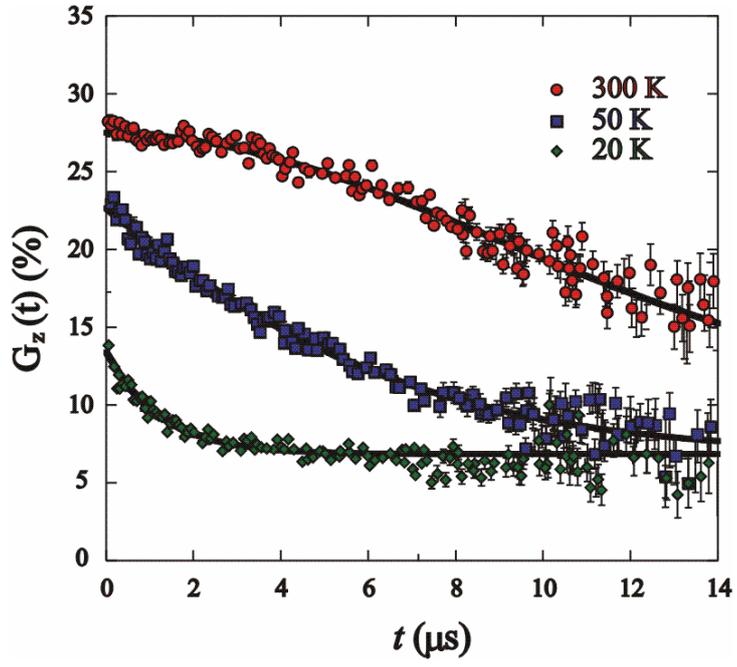

Fig 3

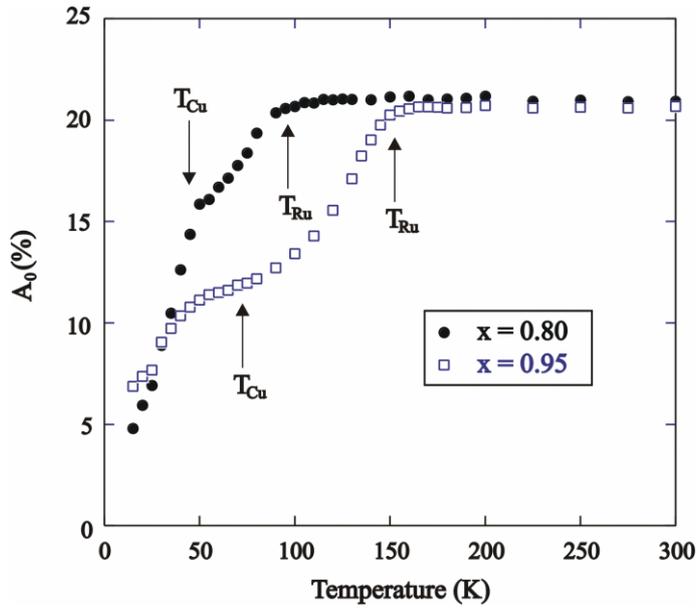

Fig. 4

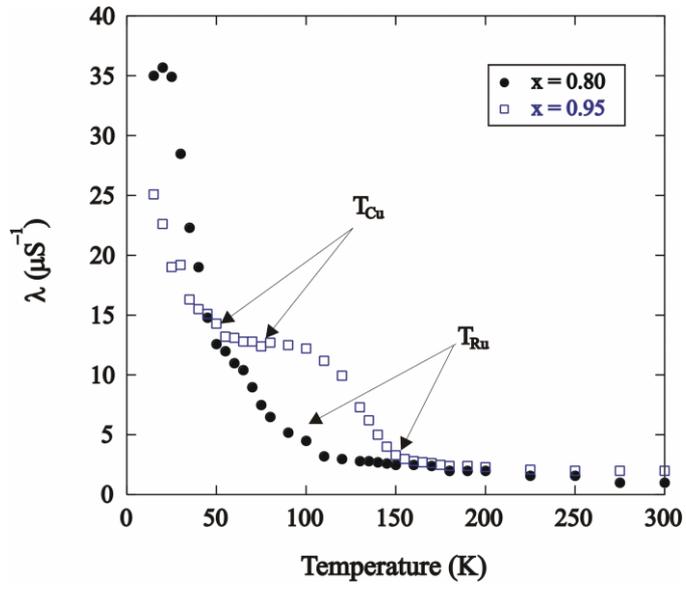

Fig 5.

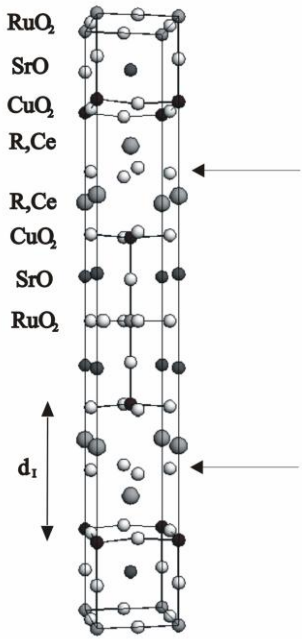

**References**

1. Bauernfeind L, Widder W and Braun H F 1996 *Physica C* **254** 151

2. Bauernfeind L, Widder W and Braun H F 1996 *J. Low Temp. Phys.* **105** 1605

3. Bernhard C, Tallon J L, Niedermayer C, Blasius T, Golnik A, Brucher E, Kremer R K, Noakes D R, Stronach C E and Ansaldo E J 1999 *Phys. Rev. B* **59** 14099

4. McLaughlin A C, Zhou W, Attfield J P, Fitch A N and Tallon J L 1999 *Phys. Rev. B* **60** 7512

5. Felner I, Asaf U, Levi Y and Millo O 1997 *Phys. Rev. B* **55**, R3374

6. Lynn J W, Keimer B, Ulrich C, Bernhard C and Tallon J L 2000 *Phys. Rev. B* **61** R14964

7. Mclaughlin A C, McAllister J A, Stout L D and Attfield J P, 2002 *Phys. Rev. B* **65** 172506

8. Mclaughlin A C, Felner I and Awana V P S 2008 *Phys. Rev. B* **78** 094501

9. Mclaughlin A C, Sher F, and Attfield J P 2005 *Nature* (London) **436** 829; Mclaughlin A C, Sher F, and Attfield J P *Nature* (London) 2005 **437** 1057

10. Mclaughlin A C, Begg L, Harrow C, Kimber S A J, Sher F and Attfield J P 2006 *J. Am. Chem. Soc.* **128** 12364

11. Awana V P S, Ansari M A, Gupta A, Saxena R B, Kishan H, Buddhikot D and Malik S K 2004 *Phys. Rev. B* **70** 104520

12. Mclaughlin A C, Begg L, McCue A J and Attfield J P 2007 *Chem. Commun.* 2273

13. Nagaev E L 1967 *JETP Lett.* **6** 18

14. Williams G V M, Jang L -Y and Liu R S 2002 *Phys. Rev. B* **65** 064508

15. Mclaughlin A C, Sher F, Kimber S A J and Attfield J P 2007 *Phys. Rev. B* **76** 094514

16. Kubo R and Toyabe T 1967 *Magnetic Resonance and Relaxation* edited by R. Blinc (North-Holland, Amsterdam) 810-823.

17. Hayano R S, Uemara Y J, Imazato J, Nishida N, Yamazaki T and Kubo R 1979 *Phys. Rev. B* **20** 850

[18] Shengelaya A, Khasanov R, Eshchenko D G, Felner I, Asaf U, Savic I M, Keller H and Muller K A 2004 *Phys. Rev. B* **69** 024517

[19] Hillier A D, Adroja D T, Giblin S R and Kockelmann W 2007 *Phys Rev. B* **76** 174439

[20] Schenck A and Gygax F N, 1995, *Handbook of Magnetic Materials* edited by K. H. J. Buschow (North-Holland, Amsterdam) **9**

[21] Bewley R I, Blundell S J, Lovett B W, Jestadt Th, Pratt F L, Chow K H, Hayes W, Battle P D, Green M A, Millburn J E, Rosseinsky M J, Spring L E and Vente J F 1999 *Phys. Rev. B* **60** 12286

[22] Watanabe I, Oki N, Adachi T, Mikuni H, Koike Y, Pratt F L and Nagamine K 2006 *Phys. Rev. B* **73** 134506

[23] Chmaissem O, Jorgensen J D, Shaked H, Dollar P and Tallon J L 2000 *Phys. Rev. B* **61** 6401